\documentclass[
 twocolumn,
 longbibliography,
superscriptaddress,
 amsmath,amssymb,
 aps,
]{revtex4}

\usepackage{graphicx}
\usepackage{dcolumn}
\usepackage{bm}

\usepackage{natbib}
\begin{document}

\preprint{APS/123-QED}

\title{Sliding or Rolling?\\Characterizing single-particle contacts}

\author{Simon Scherrer}
\affiliation{Department of Materials, ETH Zürich, Vladimir-Prelog-Weg 1-5/10, 8093, Zürich, Switzerland
}
\author{Shivaprakash N. Ramakrishna}
\affiliation{Department of Materials, ETH Zürich, Vladimir-Prelog-Weg 1-5/10, 8093, Zürich, Switzerland
}
\author{Vincent Niggel}
\affiliation{Department of Materials, ETH Zürich, Vladimir-Prelog-Weg 1-5/10, 8093, Zürich, Switzerland
}
\author{Chiao-Peng Hsu}
    \affiliation{Chair for Cellular Biophysics, Center for Functional Protein Assemblies, Center for Organoid Systems, Department of Bioscience, Technical University of Munich, Technical University of Munich School of Natural Sciences, Garching 85748, Germany}
\author{Robert W. Style}
\affiliation{Department of Materials, ETH Zürich, Vladimir-Prelog-Weg 1-5/10, 8093, Zürich, Switzerland
}
\author{Nicholas D. Spencer}
\affiliation{Department of Materials, ETH Zürich, Vladimir-Prelog-Weg 1-5/10, 8093, Zürich, Switzerland
}
\author{Lucio Isa}
 \email{lucio.isa@mat.ethz.ch}
\affiliation{Department of Materials, ETH Zürich, Vladimir-Prelog-Weg 1-5/10, 8093, Zürich, Switzerland
}

\date{\today}

\begin{abstract}
Contacts between particles in dense, sheared suspensions are believed to underpin much of their rheology. Roughness and adhesion are known to constrain the relative motion of particles, and thus globally affect the shear response, but an experimental description of how they microscopically influence the transmission of forces and relative displacements within contacts is lacking. Here we show that an innovative colloidal-probe atomic force microscopy technique allows the simultaneous measurement of normal and tangential forces exchanged between tailored surfaces and microparticles while tracking their relative sliding and rolling, unlocking the direct measurement of coefficients of rolling friction, as well as of sliding friction. We demonstrate that, in the presence of sufficient traction, particles spontaneously roll, reducing dissipation and promoting longer-lasting contacts. Conversely, when rolling is prevented, friction is greatly enhanced for rough and adhesive surfaces, while smooth particles coated by polymer brushes maintain well-lubricated contacts. We find that surface roughness induces rolling due to load-dependent asperity interlocking, leading to large off-axis particle rotations. In contrast, smooth, adhesive surfaces promote rolling along the principal axis of motion. Our results offer direct values of friction coefficients for numerical studies and an interpretation of the onset of discontinuous shear thickening based on them, opening up new ways to tailor rheology via contact engineering.
\end{abstract}

\maketitle


\section{\label{sec:level1}INTRODUCTION}

Flows of dense microscopic particle suspensions can be found in many natural and industrial settings \cite{VanDamme2018ConcreteInnovations, Blanco2019ConchingContent, Morris2020ShearPhenomena}, and have attracted much interest, as they can exhibit striking non-linear rheological responses when subjected to shear stress, $\sigma$. 
Most commonly, they exhibit shear thickening (ST), whereby $\sigma$ increases super-linearly with shear rate, $\dot{\gamma}$.
In this case, the viscosity of the suspensions increases with $\dot{\gamma}$ -- either gradually (continuous shear thickening, CST), or diverging at a critical shear rate (discontinuous shear thickening, DST).
In shear jamming (SJ), the most severe form of DST, suspensions completely solidify under shear \cite{Brown2014ShearJamming, Morris2020ShearPhenomena, Wagner2009ShearDispersions, Brown2010GeneralitySuspensions}.
ST can be advantageous, for example in impact absorption \cite{Lee2003TheFluid}, but it also causes widespread issues, in particular as it limits the processing speed of dense slurries such as concrete \cite{Ovarlez2020DensitySuspensions}. 

The onset of ST is attributed to the emergence of local constraints that restrict relative particle motion under flow, leading to enhanced dissipation \cite{Guy2018Constraint-BasedRheology}. 
As a result, any mechanism promoting inter-particle contacts will affect ST.
Indeed, recent experiments have shown that tuning particle surface coatings \cite{Fernandez2013MicroscopicSuspensions, Hsu2021ExploringParticles} or roughness \cite{Hsu2018Roughness-dependentThickening, Schroyen2019StressHairy, Comtet2017PairwiseSuspensions, AkbariFakhrabadi2021AlteringSilica, dAmbrosio2023TheSuspensions, Hsiao2017RheologicalFlow, Lootens2005DilatantParticles}, and adding short-range adhesive interactions \cite{James2019TuningRheology, Hsu2021ExploringParticles, Park2019ContactMixtures, Gauthier2023ShearCornstarch, James2018InterparticleSuspensions} yield dramatic effects on DST.

Even if recent progress has been made in identifying factors affecting ST, we currently do not have the ability to predict its onset.
Enhanced dissipation at particle contacts is typically treated as originating from a stress-activated transition from hydrodynamic to boundary lubrication, where fluid films separating the particles break down and their surfaces come into frictional and adhesive contacts \cite{Lin2015HydrodynamicSuspensions, Royer2016RheologicalSuspensions, Clavaud2017RevealingSuspensions, Morris2018Lubricated-to-frictionalSuspensions, Mari2014ShearSuspensions, Seto2013DiscontinuousSuspensions, Fernandez2013MicroscopicSuspensions, Wyart2014DiscontinuousSuspensions, James2019TuningRheology, Hsu2021ExploringParticles, Singh2020ShearFriction}. However, surface roughness coupled to hydrodynamics can also be sufficient to constrain the relative motion between particles and lead to ST \cite{Jamali2019AlternativeHydrodynamics}, making it \textit{a priori} challenging to identify the dominating cause. 
Moreover, initial DST models have only implemented sliding friction as a constraint, ignoring the possibility of rolling particle contacts, which allow relative particle motion, but with much less friction. Inspired by the rheology of dry granular suspensions \cite{Santos2020GranularFriction, Ai2011AssessmentSimulations, Estrada2011IdentificationMedia}, more recent models also explictly added rolling resistance to capture the experimentally observed onset of DST \cite{Singh2020ShearFriction, Mari2019ForceThickening, Singh2022Stress-activatedRheology, dAmbrosio2023TheSuspensions} but experimental values of rolling friction coefficients are not reported and the relative importance of sliding versus rolling contacts around multiple axes remains poorly understood. 
Finally, recent simulations suggest the importance of the dynamics of inter-particle contacts \cite{Nabizadeh2022StructureSuspensions}, in connection to the formation of a rigid percolating network associated to the onset of DST \cite{Goyal2024FlowSuspensions}. 
To understand ST, we therefore need experiments that are capable of observing and quantifying such micro-scale particle contacts.
However, this is challenging for most techniques, which cannot distinguish between sliding and rolling contacts, measure particle contact durations, nor observe multi-axis particle rotation while measuring forces.
Such techniques include methods based on optical imaging \cite{Cheng2011ImagingSuspensions}, the use of mechano-sensitive dyes \cite{Pan2015SRheology}, the rapid immobilization of sheared suspensions with inference from macroscopic rheology \cite{Pradeep2021JammingThickening} and, finally, atomic-force microscopy (AFM)-based techniques \cite{Carpick1997ScratchingMicroscopy,Fernandez2015DirectMicrospheres, Comtet2017PairwiseSuspensions}.
In the latter case, particles are immobilized on an AFM tip, only allowing for sliding contacts. However, a few studies have investigated aspects of rolling \cite{Heim1999AdhesionParticles, Bonacci2022YieldSuspensions, Schiwek2016EvidenceEnvironment, Ding2007RollingMeasurements}, typically by using sharp AFM tips to manipulate individual microparticles \cite{Sitti2004AtomicCharacterization, Sumer2008RollingPushing}.
 
Here, we present an alternative experimental approach, based on colloidal-probe AFM (CP-AFM).
Our method can characterize nano-scale contacts of free microparticles in a liquid while measuring normal and friction forces at the contacts, and simultaneously imaging both rotation and translational particle motions.
By studying particles with defined surface properties we measure both sliding and rolling friction coefficients and show that contacts switch from sliding to rolling if sufficient traction is present — due to the interlocking of surface asperities or adhesion.
This transition to rolling is accompanied by a significant increase in the average duration of particle contacts.
Although we might expect that roughness and adhesion play similar roles in ST, they cause types of particle motion that are qualitatively very different.
While smooth, adhesive contacts induce particle rotation aligned with the main shear direction, rough contacts have extremely large off-axis angular displacements, caused by interlocking of random surface asperities.
Our results have strong implications for force transmission in the DST regime and shed light on the microscopic processes that underlie the DST onset. By examining rheology data we show that adhesion reduces the critical stress for the onset of DST, which is further reduced if the particles are also rough. Such particles, if rolling is prevented within a contact network, experience a significant increase in effective friction, which in turn leads to a stronger DST, i.e. a larger viscosity jump at lower volume fractions. Furthermore, our technique offers an approach for quantifying microscopic rolling friction coefficients, which can be used as a basis for developing accurate predictive models of ST.

\section{\label{sec:level2}RESULTS}
\subsection*{Measurements with free colloidal probes}

\begin{figure*}[ht]%
\centering
\includegraphics{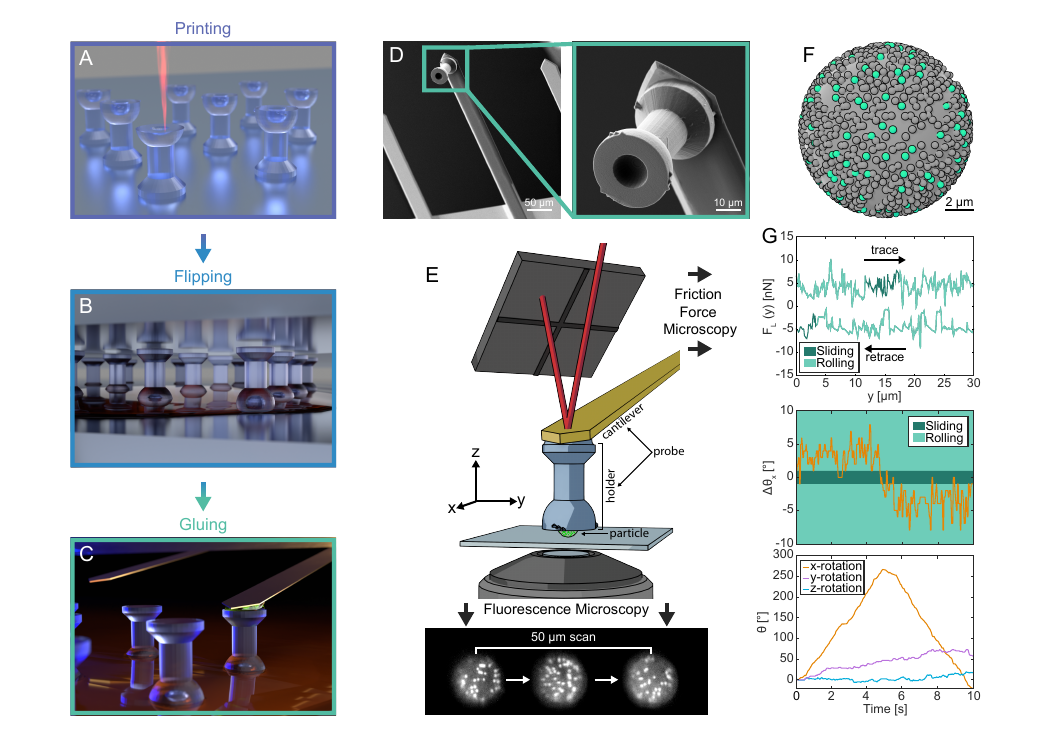}
\caption{Schematic overview of the fabrication of the probe and the experimental workflow. A) Printing: 3D 2PP-DLW of an array of holders. B) Flipping: The holder array is transferred onto a glass slide presenting a sacrificial glucose layer. C) Gluing: A single holder is glued to a tipless AFM cantilever with UV glue and released by dissolving the sacrificial layer. D) SEM scan of the assembled probe, consisting of the AFM cantilever and the particle holder. E) Experimental setup (not to scale): A captured particle is moved across a transparent substrate along the y-axis, measuring normal and lateral forces with an AFM while collecting a time series of fluorescence images from below using an inverted microscope. F) Schematic of a typical silica raspberry (RB) particle consisting of a 12 $\mu$m core decorated with nanoparticles of which a fraction is fluorescent. G) A representative friction loop, showing the lateral force $F_L$ as a function of scanning distance $y$, obtained from scanning a RB particle on a rough substrate (top). The particle motion (rolling/sliding) is extracted from the image sequence and rolling is identified if the angular displacement exceeds $\pm$ 1\textdegree\ between subsequent images (middle). The total 3D angular displacements during a friction loop are measured to extract the ratio of sliding to rolling (bottom).}
\label{fig1}
\end{figure*}

Our approach combines a novel CP-AFM holder as the force-sensing unit with an inverted epi-fluorescence widefield microscope \cite{Scherrer2024MeasuringNanoscale}. The instrument allows the simultaneous analysis of the forces acting at contact and the determination of the relative motion between a particle and a flat countersurface as schematically shown in Fig. \ref{fig1}. 
We fabricate custom-made particle holders by means of 2-photon polymerization direct laser writing (2PP-DLW) and affix them to the end of an AFM cantilever. The holders contain a hemispherical cavity that is slightly larger than the diameter of our particles of interest so that particles trapped inside are free to rotate (see Figs. \ref{fig1}A-D, Fig. S1 and \textit{Materials and Methods} for an overview of the fabrication process).  

Using this setup, we capture particles on a flat countersurface (all immersed in an aqueous buffer, see video S1), and translate them laterally by moving the holder, while measuring lateral and normal forces with the AFM (see videos S2-S11).
The force measurement proceeds as in conventional CP-AFM, where the raw voltage signals measured by the AFM photodetectors during lateral scanning are converted to forces after calibration. Our system allows for the use of established calibration methods, i.e. the wedge-calibration method for lateral-force calibration \cite{Varenberg2003AnMicroscopy} and Sader's method for cantilever-spring-constant calibration \cite{Sader1999CalibrationCantilevers}, which we adapted to the new geometry of the probes (see also Fig. S2 and \textit{Materials and Methods}).
Simultaneously, we quantify the translation and rotation of the particles by imaging them with an inverted epi-fluorescence microscope (Fig. \ref{fig1}E). 
In the experiments presented here, we use particles with tailored roughness and surface coatings. Rough particles, which we call raspberry-like particles (RB particles), are made of a 12 $\mu$m silica core and are decorated with silica nanoparticles of either 100 or 300 nm diameter to control their surface roughness. A fraction of the nanoparticles contains a fluorescent dye, enabling them to act as tracers for rotation tracking \cite{Ilhan2020SphericalDimensions, Liu2016MeasuringMicroscopy}. Smooth particles comprise the same 12 $\mu$m silica cores, but are covered with fluorescent quantum dots (QDs) embedded in a smooth silica layer, leading to a heterogeneous fluorescence, which also allows rotation tracking (see also Fig. \ref{fig1}F, Fig. S3 and \textit{Materials and Methods}). In general, this hetero-aggregation approach to assemble particles has the benefit of a high degree of control over the surface properties, which can be tuned independently of the fluorescent markers, and of providing a chemically homogeneous silica surface for easy modification, e.g. by polymeric coatings, as described later \cite{Zanini2017FabricationHeteroaggregation, Hsu2018Roughness-dependentThickening, Hsu2021ExploringParticles}. 
From the changes of the fluorescence signal during scanning, we can thus measure relative particle motion, distinguish rolling from sliding, and quantify particle 3D rotation via an image-correlation algorithm  \cite{Niggel20233-DSurfacesb}. (Details on the working principle of the algorithm are given in Fig. S4 and \textit{Materials and Methods}.)

An example of results from a typical experiment is shown in Fig. \ref{fig1}G.
We move the cantilever laterally in the $y-$direction (see Fig. \ref{fig1}E for axis definitions) while recording friction loops (top) of the lateral force signal, $F_L$, versus scanning distance $y$ at a fixed normal load during trace and retrace.
At the same time, we measure particle rotation around the $x$, $y$, and $z$ axes, and plot the change in angle between frames (middle -- here showing rotation around the $x-$axis) to correlate the local lateral force with the corresponding angular motion.
Additionally, we integrate these angular differences to obtain the total rotation of the particle during a friction loop (bottom). This rotation can then be compared to the lateral displacement of the probe to extract the amount of slip, and thus the traction of the RB particle on the substrate. As might be anticipated, the angular displacements are predominantly around the $x-$axis. Later, we will show that this is not always the case.

We can use this apparatus to measure sliding- and rolling-friction coefficients between the particle and substrate. When a colloidal particle slides across a substrate, the measured lateral force $F_L$ originates exclusively from the particle/substrate interaction during sliding, thereby providing direct access to the friction force at the contact point.
In this case, the sliding friction coefficient $\mu$ can be calculated following a modified version of Amonton's law \cite{Gao2004FrictionalScale} 
\begin{equation}
    F_L = \mu F_N + F_L^0,
    \label{amonton}
\end{equation}
where $F_N$ is the applied normal load and $F_L^0$ is the friction due to the adhesion force.

\begin{figure}[ht]%
\centering
\includegraphics{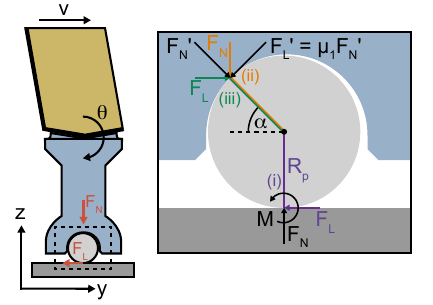}
\caption{Schematic of a free colloidal probe inside the holder during lateral translation. The zoom-in shows the relevant forces and moments that the holder and the substrate exert on the particle.}
\label{fig2}
\end{figure}

When a particle rolls against a substrate, the substrate exerts a moment $M$ on the particle -- which opposes the rolling motion (see Fig. \ref{fig2}). This rolling-friction moment is also expected to obey the modified version of Amonton's Law:
\begin{equation}
\frac{M}{R_p} = \frac{\eta}{R_p} F_N+M_0, \label{rolling}
\end{equation}
which allows us to obtain the dimensionless rolling-friction parameter, $\eta/R_p$, which is analogous to the sliding friction coefficient $\mu$. Here, $R_p$ is the particle radius, $F_N$ is again the applied normal force, and $M_0$ is the adhesive rolling-friction moment. It will be demonstrated that $M_0$ is too small to be measurable by our technique and thus is neglected in the following analysis.

To obtain $M$, we identify the moment balance on a steadily rolling particle (Fig. \ref{fig2}). This gives that:
\begin{equation}
\frac{M}{R_p} = F_L - F_N \cos\alpha +F_L \sin\alpha, \label{M}
\end{equation} where $F_L$ is again the lateral force exerted on the particle by the AFM.
Here, the first term on the right-hand side comes from the torque due to the particle/substrate friction force, and the last two terms represent the torque from the force at the particle/holder contact. $\alpha$ is the polar angle of the particle/holder contact point relative to the plane of the substrate (Fig. \ref{fig2}), which needs to be determined to obtain $M$.
When the particle rolls across the substrate, it slides inside the holder. The latter contact point thus obeys Amonton's law as: $$F'_L= \mu_1 F'_N$$ (this simple relation holds as there is no adhesion between the particle and the holder, see Fig. S5) which can be rewritten as
\begin{equation}
F_L\cos\alpha+F_N\sin\alpha=\mu_1(F_N\cos\alpha-F_L\sin\alpha),
\label{particle/holder}
\end{equation}
from which we can obtain $\alpha$, if we know the particle/holder friction coefficient $\mu_1$.  
The value of $\mu_1$ can be directly accessed from friction loops at the onset of rolling (here, taken as loops where the percentage of time spent rolling versus sliding is between 30 and 70\%), where $M \approx 0$, so that (from equation \ref{M}) $F_L(1+\sin\alpha)=F_N\cos\alpha$. Inserting this expression into equation (\ref{particle/holder}), we obtain the constant $\mu_1$. Because this value is specific to the particle/holder pair and does not depend on the substrate, it can then be used with the results above to calculate $\eta/R_p$ for all rolling datasets against different substrates.

\subsection*{Impact of surface roughness}\label{rough}

Among the particle characteristics previously mentioned, surface roughness has been clearly shown to promote the onset of DST  \cite{Hsu2018Roughness-dependentThickening, Hsiao2017RheologicalFlow, Jamali2019AlternativeHydrodynamics, Lootens2005DilatantParticles}. CP-AFM studies have shown that roughness constrains relative particle motion by the interlocking of surface asperities, but it remains unclear how it affects contact forces if particles are free to roll rather than being forced to slide past each other \cite{Hsu2018Roughness-dependentThickening}. 

We therefore start our measurements with RB particles with 300 nm asperities and translate them across surfaces with well-defined roughness.
This roughness is controlled by attaching nanoparticles to glass slides with two separate strategies. In the first strategy, we create densely packed layers of monodisperse nanoparticles, with diameters of 100, 200, 300, 400 or 500 nm. This process gives substrate roughnesses with a range of different lengthscales relative to the RB particle roughness. In the second strategy, we attach 50 nm nanoparticles to the glass using different adsorption times. This changes the number density of adsorbed nanoparticles.
In all cases, surface roughness is characterized via the parameter $h/d$, defined as the ratio between the average asperity height, $h$, and the average inter-asperity distance, $d$.
Using the first strategy we fabricated substrates with $h/d$ values between 0.26 and 0.57 (see Fig. S6), while the second strategy produced substrates with $h/d$ values between 0.06 and 0.42 (see Fig. S7). Further details on the roughness characterization of the particles and substrates can be found in Fig. S6, Fig. S7, Fig. S8, and in \textit{Materials and Methods}.

\begin{figure*}[ht]%
\centering
\includegraphics[width=0.99\textwidth]{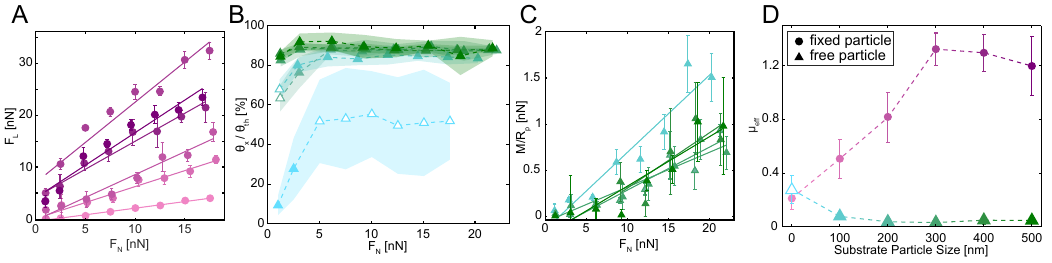}
\caption{Friction and corresponding motion of silica raspberry (RB) particles (12 $\mu$m core with 300 nm nanoparticle asperities) on complementary rough substrates. A) Lateral force $F_L$ versus normal force $F_N$ for fixed RB particles against substrates with asperities of varying size (the size can be read off the $x$-axis of panel D for correspondingly colored points). B) Rotation of free RB particles on substrates with asperities of varying size, expressed as a percentage of total angular displacement along the $x$-axis $\theta_x$ during a friction loop compared to that of pure rolling without slip $\theta_\mathrm{th}$, as a function of the applied normal force. Substrate particle size can be read off the $x$-axis of panel D for correspondingly colored points, and empty symbols represent data points with $\theta_x$/$\theta_\mathrm{th}$ = 30 - 70 \% from which $\mu_1$ was extracted. C) Calculated rolling resistance $M/R_p$ versus normal force $F_N$ for free RB particles with $\theta_x$/$\theta_\mathrm{th} > 50 \%$ on substrates with asperities of varying size (same data sets/color as in (C)). D) Evolution of the effective friction coefficient $\mu_\mathrm{eff}$ for fixed ($\mu_\mathrm{eff} = \mu$) and free ($\mu_\mathrm{eff} \approx\eta/R_p$) RB particles on rough substrates with increasing asperity/substrate-adsorbed particle size. Empty symbols denote data points for which $\theta_x$/$\theta_\mathrm{th} < 50 \%$, such that the holder contribution cannot be fully removed.}
\label{fig3}
\end{figure*}

Fig. \ref{fig3} shows the frictional behavior of fixed (i.e. rigidly bonded to an AFM cantilever) and free-to-rotate particles when translated across rough surfaces. 
Each measurement is the average of several friction loops (see Fig. S9).
We start with fixed particles in Fig. \ref{fig3}A, where we show $F_L$ as a function of applied normal force, for particles translating across rough surfaces of a range of different asperity sizes. 
These datasets are all approximately linear -- as predicted by Amonton's law -- so the slope of the data gives the friction coefficients, $\mu$, which are shown as the pink-purple data points in Fig. \ref{fig3}D.
We find that $\mu$ is relatively low on smooth substrates, and increases with substrate asperity size up to asperities of 300 nm in diameter, whereupon it peaks and slowly starts to decrease.
This peak occurs when the particle roughness asperities are the same size as the substrate roughness asperities.
At this point, we expect maximal particle/substrate asperity interlocking, and thus a corresponding maximal $\mu$.
Note that a clear signal of asperity interlocking is seen by the presence of large fluctuations in the lateral force signal on rough surfaces (Fig. S9).

Figs. \ref{fig3}B,C show the results for particles that are free to rotate in the holder, as they translate across the same rough substrates from above.
Fig. \ref{fig3}B shows the rolling percentage $\theta_x/\theta_\mathrm{th}$ during a friction loop as a function of the applied normal force, which represents the proportion of time spent rolling versus sliding (the corresponding data for substrates with an increasing number density of asperities are found in Fig. S10). Here, $\theta_x$ is the measured integrated angular displacement along the principal rotation axis, while $\theta_\mathrm{th} = s/R_p$ is the angular displacement that would occur during pure rolling motion, where $s$ is the total scanning distance during one loop.
On a smooth substrate (light blue), the particles slide at very low $F_N$ and increase their rolling percentage to $\sim 50\%$ as $F_N$ increases.
On the rough substrates, particles predominantly roll, with only small reductions in time spent rolling at the lowest applied $F_N$ (see also additional data in Fig. S11 and video S2).
This rolling is caused by the interlocking of the asperities between the particle and substrate, which prevents relative motion at the contact point.
By contrast, on smooth surfaces, and at low $F_N$, we hypothesize that the ability to interlock is reduced due to the presence of a hydration-lubrication layer that prevents asperity-asperity contacts \cite{Gaisinskaya2012HydrationParadigm}.
Then, particles spend a significant fraction of the time sliding, apart from discrete events where they catch large local defects/asperities on the substrate, which induces temporary rolling (videos S3,S4).
This erratic switching between rolling and sliding is indicated by the large standard deviation of the rolling percentage in Fig. \ref{fig3}B. 

The frictional behavior of free particles is very different from that of fixed particles (Figs. \ref{fig3}C,D).
For rolling particles ($\theta_x/\theta_\mathrm{th}>50\%$), we use equations (\ref{M},\ref{particle/holder}) to extract the rolling-friction moment, $M$ as a function of $F_N$, as shown in Fig. \ref{fig3}C (see further details in Fig. S12).
As predicted, $M$ obeys the Amonton-like law (\ref{rolling}), with $M_0\approx 0$. The slopes of these curves then give the rolling-friction coefficients $\eta/R_p$, which are shown as a function of surface roughness in Fig. \ref{fig3}D as the triangular points, i.e. $\mu_{\mathrm{eff}} \approx \eta/R_p$.
For all rough substrates $\eta/R_p$ is much smaller than $\mu$. Moreover, unlike the sliding-friction coefficient $\mu$, $\eta/R_p$ barely changes with surface roughness, reflecting the ease with which rough particles can roll across rough surfaces (note that the value of the $\eta/R_p$ for the smooth substrate should be taken with caution, as the particle does not undergo pure rolling, but performs a $\sim$50-50 mixture of sliding and rolling).

It is important to point out that, for all these experiments, the asperities on the particles and substrates should not significantly deform.
Thus, it is predominantly interlocking that controls frictional behavior and not changes in the number of contacting asperities as $F_N$ increases.
We verified this by modeling the contact between our silica RB particles and a smooth silica substrate following a numerical continuum mechanics approach proposed by Röttger et al. \cite{Rottger2022Contact.engineeringCreateScales}. The results can be seen in Fig. S13 and show that, in our experiments, the contact area of a single nanoparticle of 100-500 nm diameter with a flat, silica substrate did not exceed 120 nm$^{2}$, corresponding to negligible Hertzian deformation. 

\subsection*{Impact of adhesion}\label{adhesive}

Beyond roughness, a second, key surface property, which affects contact interactions and thus strongly impacts DST, is adhesion \cite{James2019TuningRheology, James2018InterparticleSuspensions, Hsu2021ExploringParticles}.
To this end, we investigate a modified version of our smooth and rough particle system that allows us to reversibly tune particle/substrate adhesion with temperature (see Fig. \ref{fig3}, Fig. S14).
In particular, we functionalize smooth and rough silica particles and complementary glass substrates with a poly(N-isopropylacrylamide) (PNIPAM) brush via surface-initiated atom transfer radical polymerization \cite{Matyjaszewski1999PolymersInitiator, Mandal2019TuningSolvents}. Both particle types comprise a 12 $\mu$m core, decorated with either fluorescent quantum dots embedded in a silica shell (smooth) or 100 nm nanoparticles, of which a fraction is fluorescent (rough) (see \textit{Materials and Methods}). PNIPAM is a thermo-responsive polymer, which transitions from a hydrophilic, swollen state to a hydrophobic, collapsed state at a lower critical solution temperature (LCST) of around 30 \textdegree C in aqueous environments \cite{Halperin2015PolyN-isopropylacrylamideResearch}. 
Below the LCST, the polymer chains extend into the water, partially or fully masking the underlying surface topography and acting as an efficient lubricating layer.
Above the LCST, the collapsed chains not only expose the underlying roughness, but PNIPAM-functionalized surfaces also experience pronounced adhesion due to hydrophobic interactions \cite{Greve1999VariablePolymerizations, Hsu2021ExploringParticles}.

\begin{figure*}[ht]%
\centering
\includegraphics[width=1\textwidth]{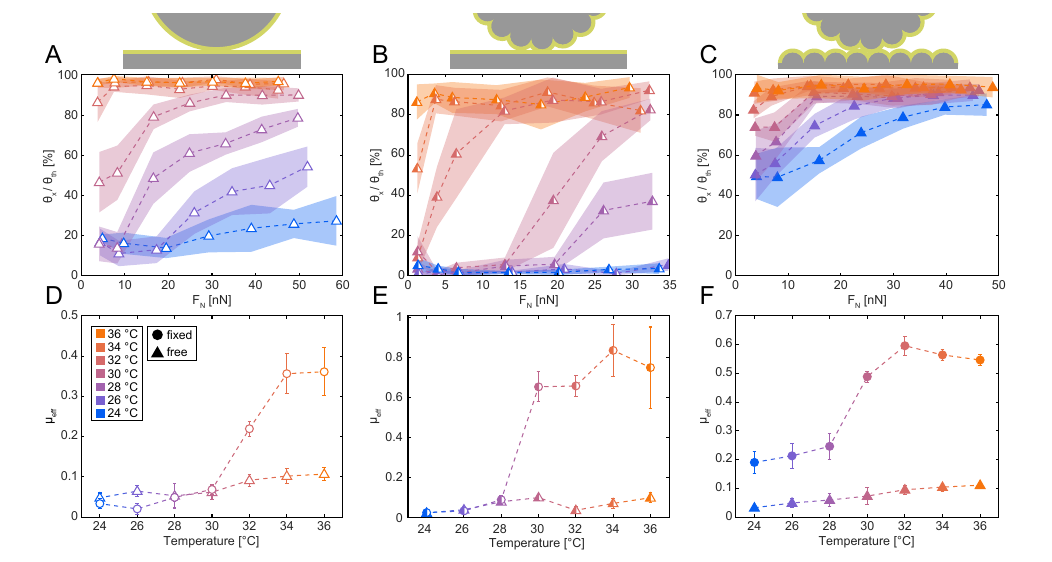}
\caption{The rolling percentage, expressed as a percentage of total angular displacement along the $x$-axis $\theta_x$ during a friction loop compared to that of pure rolling without slip $\theta_\mathrm{th}$, as a function of the applied normal force for the temperature range between 24 \textdegree C and 36 \textdegree C of A) smooth PNIPAM-coated particles on a smooth PNIPAM-coated substrate, B) rough PNIPAM-coated RB particles (12 $\mu$m core with 100 nm nanoparticle asperities) on a smooth PNIPAM-coated substrate, C) rough PNIPAM-coated RB particles (12 $\mu$m core with 100 nm nanoparticle asperities) on a rough PNIPAM-coated substrate (100 nm nanoparticle asperities). Friction coefficient ($\mu_{\mathrm{eff}}$) for fixed ($\mu$) and free ($\mu$ if $\theta_x/\theta_\mathrm{th}<50\%$, $\eta/R_p$ if $\theta_x/\theta_\mathrm{th}>50\%$) for the temperature range between 24 \textdegree C and 36 \textdegree C of D) smooth PNIPAM-coated particles on a smooth PNIPAM-coated substrate, E) rough PNIPAM-coated RB particles (12 $\mu$m core with 100 nm nanoparticle asperities) on a smooth PNIPAM-coated substrate, F) rough PNIPAM-coated RB particles (12 $\mu$m core with 100 nm nanoparticle asperities) on a rough PNIPAM-coated substrate (100 nm nanoparticle asperities).}
\label{fig4}
\end{figure*}

Our PNIPAM-coated surfaces also exhibit the reported temperature-dependent adhesion, which we characterize using fixed-particle CP-AFM.
For smooth substrates and particles, $F_\mathrm{adh}\approx 0$ nN at 24$^\circ$C (in the hydrophilic state), but increases to 10 nN at 36$^\circ$C (in the hydrophobic state). Between these two limits, $F_\mathrm{adh}$ increases with increasing temperature, displaying a sharp jump between 30 and 32 \textdegree C, i.e., across the LCST (Fig. S14). Smooth-substrate/rough-particle adhesion is quantitatively similar, while rough-substrate/rough-particle adhesion is 6-7 times smaller due to a reduction of the effective contact area that results between asperities.

Adhesion profoundly impacts the motion of free particles.
This is demonstrated in Figs. \ref{fig4}A,B,C, which show the percentage of rolling for smooth-particle/smooth-substrate, rough-particle/smooth-substrate, and rough-particle/rough-substrate contacts respectively (see also Videos S7-10).
At low temperatures, we see that a large fraction of time is spent sliding -- especially when one of the surfaces is smooth, due to the lubrication offered by the swollen PNIPAM brushes (see also Fig. S15A).
When the temperature increases above the PNIPAM LCST, we see a switch to almost complete rolling, driven by the large increase in adhesion.
At intermediate temperatures, we see an increase in rolling fraction with $F_N$, similar to the behavior seen earlier in rough-particle/rough-substrate friction.

Adhesion also has a strong influence on particle substrate friction.
The circular data points in Figs. \ref{fig4}D,E,F show the coefficient of friction for fixed particles moving across substrates.
For all particle/substrate combinations, friction comes exclusively from sliding, and $\mu$ is small at low temperatures but jumps up by several times as $T$ crosses the LCST, as previously reported \cite{Hsu2021ExploringParticles}.

By contrast, the triangular points in Figs. \ref{fig4}D,E,F show the effective friction coefficients for particles that are free to roll.
If $\theta_x/\theta_\mathrm{th} <50\%$, the particles spend most of their time sliding and $\mu_\mathrm{eff} \approx \mu$. Conversely, if $\theta_x/\theta_\mathrm{th}  >50\%$, particles predominately roll and $\mu_\mathrm{eff} \approx \eta/R_p$, which we calculate following the procedure described earlier.
For all the particle/substrate combinations, the effective friction only shows a little increase with temperature. This increase is substantially smaller than the corresponding increase in $\mu$, and there is no jump at the LCST. Rather, the free particles transition from predominantly sliding to rolling by crossing the LCST. In other words, free particles, if allowed, spontaneously start rolling in the presence of adhesion. The measured values of $\eta/R_p$ are $\lesssim 0.1$, which falls within the range of the values used in numerical simulations \cite{Singh2020ShearFriction}. 
We hypothesize that once rolling takes place, the rolling friction coefficient is less influenced by variations in adhesion than sliding friction. In fact, adhesion acts as an additional normal force at contact and it induces extra dissipation associated to breaking and forming bonds within the moving contact. In the case of sliding friction, the extra dissipation 
is connected to hysteresis between breaking and forming those bonds over the whole contact area. Conversely, for pure rolling, only the bonds at the back of the particle are broken and dissipation is related to hysteresis between breaking and forming bonds along the outer edge of the contact region, i.e. the contact perimeter. We thus hypothesize that the adhesive contribution against sliding scales with contact area, while it scales with contact perimeter for rolling, leading to smaller detectable variations.

The data for rough-particle/rough-substrate in Fig. \ref{fig4}F allows us to investigate the combined effect of adhesion and roughness upon rolling and sliding friction, i.e. for 100 nm asperities on both particle and substrate (light green) in Fig. \ref{fig3}B-D.
Rough PNIPAM-coated particles at 24 $^\circ C$, i.e. with no adhesion, have a lower propensity to roll than uncoated particles at all values of $F_N$, as it may be expected because the PNIPAM brush will lubricate the surfaces, making it easier for them to slide past each other.
Above around 30 \textdegree C, where the PNIPAM chains collapse to a thin adhesive layer, the coated and uncoated particles/substrates have a very similar topography and the observed rotation matches the behavior of the uncoated system well. Above the LCST, the particles almost exclusively roll, due to the combined effect of traction obtained from roughness-induced interlocking and adhesion.
The sliding friction coefficient for rough, coated particles at low temperature is lower than for uncoated particles of the same roughness, see  Fig. \ref{fig3}D but becomes higher if adhesion is turned on.
Conversely, for free particles, the rolling resistance $\eta/R_p$ appears to be only weakly affected by the polymeric coating both with and without adhesion. Finally, Fig. \ref{fig4}B and E, report the data for the mixed case of rough, PNIPAM-coated particles against a smooth surface, which exhibit an intermediate behavior between the smooth and rough cases.

\subsection*{Impact on dense suspensions}

\begin{figure}[ht]%
\centering
\includegraphics{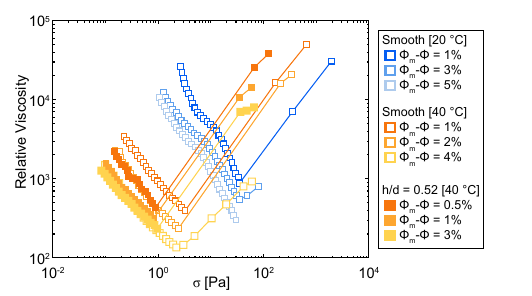}
\caption{A) Relative viscosity as a function of shear stress $\sigma$ for different particle systems, and for various volume fractions $\Phi$ relative to the maximum packing fraction $\Phi_m$. The empty symbols refer to PNIPAM-coated smooth particles without (20 \textdegree C) and with adhesion (40 \textdegree C), while the filled symbols refer to rough, adhesive RB particles with $h/d = 0.52$ (40 \textdegree C). Data adapted from  \cite{Hsu2021ExploringParticles}.}
\label{fig5}
\end{figure}

The measurement of sliding and rolling friction coefficients allows us to reexamine the ST rheology of analogous systems of particles previously reported in the literature \cite{Hsu2021ExploringParticles}. 
Fig. \ref{fig5} shows the relative viscosity of different systems of particles (i.e. defined as the suspension’s viscosity normalized by the viscosity of the medium at the corresponding temperature), as a function of shear stress $\sigma$ for different volume fractions $\Phi$, i.e. defined as relative to the maximum packing fraction of the various suspensions – $\Phi_m$.
The empty blue data points refer to smooth PNIPAM-coated silica particles at 20$^\circ$C (i.e. without any adhesion), the empty yellow-orange points refer to the same particles but in the presence of adhesion (i.e. at 40$^\circ$C), and the filled yellow/orange points correspond to the adhesive and rough RB-particles with $h/d = 0.53$.

For all these systems, we observe that the DST onset, here visible as a sharp viscosity increase as a function of $\sigma$, takes place at a system-dependent shear stress, which does not depend on volume fraction, as previously reported in the literature \cite{Mari2014ShearSuspensions}. However, this critical stress is reduced in the presence of adhesion. This fact indicates that adhesion acts as an additional normal force at contact, shifting the critical stress for the onset of frictional contacts to a lower value. In particular, we observe that the critical stress is reduced by more than one order of magnitude for PNIPAM-coated smooth particles upon turning adhesion on. The value is further reduced for rough adhesive particles. It is thus easier for adhesive particles to enter into frictional contacts than for non-adhesive ones, for which contacts remain lubricated up to higher stresses. i.e. the presence of surface asperities further facilitates frictional contacts at even lower stresses.

The friction coefficients we measured not only affect the values of $\Phi_m$ \cite{Morris2020ShearPhenomena} but also affect the viscosity jump across the DST transition when the particles are in boundary contacts. The data reported in Fig. \ref{fig4}A show that for smooth PNIPAM-coated particles at low temperatures, i.e. with no adhesion, the effective friction coefficient does not change, irrespective of whether the particles are free to roll at contact or not. Conversely, in the presence of adhesion,  the same particles experience a large jump in effective friction if rolling is constrained. Similarly, this is also seen for the rough PNIPAM-coated particles in Fig. \ref{fig4}C.
This is a crucial difference, which we hypothesize strongly affects the viscosity jump across the ST transition. Before DST, particles will roll or slide past each other, with whichever mechanism creates the lower dissipation. However, upon DST, we hypothesize that when particles come into contact within a network (local or system-spanning), they are prevented from freely rolling relative to one another by multiple contacts with their neighbors. They are thus forced to slide past each other under shear. Upon sliding, adhesive and rough particles experience much larger friction than smooth, non-adhesive ones, as seen in Fig. \ref{fig4}, leading to higher dissipation.

\subsection*{Analysis of relative motion at contact}
We have seen how both roughness and adhesion can trigger particle rolling instead of sliding, accompanied by an associated drop in friction, or conversely how blocking rolling leads to a jump up in effective friction.
While both parameters lead to qualitatively similar behavior, important differences emerge when examining fluctuations in the friction force, and when analyzing how the particles rotate against different substrates.

\begin{figure}[ht]%
\centering
\includegraphics{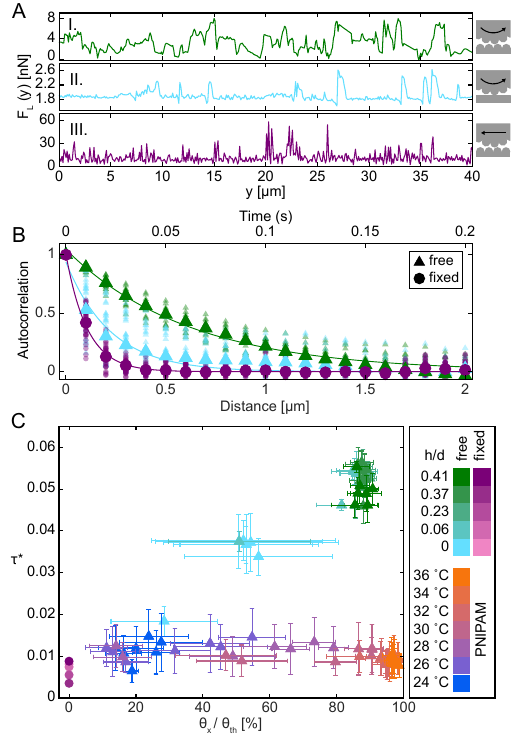}
\caption{Evaluation of the contact time via autocorrelation analysis of the lateral force signal. A) Representative lateral force signals of a RB particle (12 $\mu$m core decorated with 300 nm nanoparticle asperities) that is free to rotate on: (I.) a rough substrate with 50 nm silica nanoparticle asperities ($h/d$ = 0.41) at an applied normal force of 17 nN, (II.) a smooth substrate (h/d = 0) at an applied normal force of 1 nN, (III.) a rough substrate of 50 nm silica nanoparticles ($h/d$ = 0.41) at an applied normal force of 17 nN. B) Normalized autocorrelation function of the lateral force signal from A). C) Normalized contact duration $\tau^*$ as a function of the percentage of time spent rolling instead of sliding.}
\label{fig6}
\end{figure}

A detailed insight into the lateral force during relative motion is possible through an analysis of the force fluctuations during friction loops, of which we calculate the time autocorrelation about the average force.
For example, Fig. \ref{fig6}A shows representative lateral force signals for free (I. and II.) and fixed (III.) particles moving against a rough($h/d=0.41$, I. and III.) and a smooth substrate (II.) at an applied normal force of 1 and 17 nN respectively, i.e., taken from the experiments reported in Fig. S10.
The corresponding autocorrelations are shown in Fig. \ref{fig6}B. 
We see that the autocorrelation decays much more slowly for free particles than for fixed ones, indicating that force correlations persist much longer for free particles, and suggesting a stark difference in contact duration.
The difference can be quantified by fitting the autocorrelations to exponential functions: $Ae^{-x/\lambda}$, to extract a decay length, $\lambda$, or a decay time, $\tau=\lambda/v$, with $v$ indicating the relative speed at contact and corresponding to the cantilever scanning speed.
Representing the case of two suspended particles in a shear flow, the contact decay time is naturally compared to the characteristic time scale for the flow, i.e., the inverse of the characteristic shear rate $\dot{\gamma}=v/2R$.
We can then define the normalized contact duration, $\tau^*=\tau\dot{\gamma}=\lambda/2R$ for both free and fixed particles on substrates with a range of roughnesses, and plot it against the particle rotation in Fig. \ref{fig6}C.
We observe that the average contact duration on rough surfaces is much shorter for translating fixed particles (circles, see also Fig. S16A) than for free rolling particles (triangles).
In the former case, there is an upper bound on $\tau$ given by asperity radius divided by translation velocity, which implies that $\lambda \simeq h/2$ for our spherical asperities, i.e., $\tau^* \simeq 0.01$ as shown in Fig. \ref{fig4}C. In other words, in the case of pure sliding, interlocking asperities only stay in contact for the time required to slide over the asperity, during which the lateral force overshoots and rapidly drops after the contact is broken. 
However, the situation is different in the case of pure rolling. Here, we can assume that the upper bound for the contact time between asperities is given by the time it takes for the two spherical asperities to roll over each other, i.e., over a distance $\lambda \simeq \pi h$ -- or half the asperity equatorial perimeter. Correspondingly, $\tau^* \simeq 0.075$ for 300nm asperities and RB particles with a 12 $\mu$m diameter. This estimation correlates very well with the data reported in Fig. \ref{fig4}C for pure rolling, i.e., where $\theta_{x}/\theta_\mathrm{th} = 100$\%. Partial rolling gives intermediate average contact durations due to the occurrence of both rolling and sliding. 

Smooth PNIPAM-functionalized particles moving against a flat, PNIPAM-functionalized substrate show a very different behavior. Here, the differences between free and fixed particles are much less pronounced (Fig. \ref{fig6}C and Fig. S16B), and $\tau^*$ does not depend on the percentage of rolling. These differences can be ascribed to the different mechanisms underpinning contact mechanics below and above the PNIPAM LCST. Small values for $\tau^*$ in the non-adhesive regime below 30 \textdegree C suggest short interaction times due to the lubricated nature of the contact, where fluctuations in the lateral force may originate from heterogeneities in the polymer brushes. Rolling above the LCST is due to the traction generated by adhesion, as opposed to the case of geometrical asperity interlocking for rough surfaces, as just described above. Again here, we expect that contacts are dominated by the formation and breaking of hydrophobic interactions of the PNIPAM molecules and the homogeneity of the collapsed polymer brushes as the particles roll over the flat glass substrate, which appears to be independent of rolling percentage.

\begin{figure}[ht]%
\centering
\includegraphics{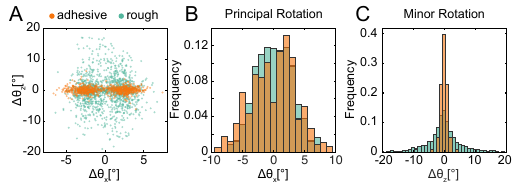}
\caption{Particle rotation around the principal (x-) and minor (z-) axis. A) Scatter plot of $\Delta\Theta_x$ against $\Delta\Theta_z$ of a rough surface (RB particle on a substrate with 50 nm silica nanoparticles $h/d = 0.06$ at 17 nN applied normal force) and an adhesive surface (PNIPAM functionalized smooth particle and substrate at 36 \textdegree C at 18 nN applied normal force). B) Histogram of the principal rotation of the particles ($\Delta\Theta_x$) of the rough and adhesive systems in A during displacement in the positive and negative y-direction. C) Histogram of the minor rotation of the particles ($\Delta\Theta_z$) of the rough and adhesive systems in A during scanning in the positive and negative y-direction.}
\label{fig7}
\end{figure}

Roughness-induced rolling and adhesion-induced rolling are also very different in how they impart rotation to particles.
During the rolling friction measurement, a free particle is pushed along the $y-$axis, such that the principal rotation occurs naturally around the perpendicular $x-$axis.
Nevertheless, the particle is not restricted to rotations along this one axis, and rotations around the $y-$ and $z-$axes are also possible (the latter corresponds to spinning, see video S11). However, their occurrence depends on the nature of the contact.
Fig. \ref{fig7}A showcases how spinning around the $z-$axis can be surprisingly large, especially for free particles on a rough surface, here for example with 50 nm asperities ($h/d = 0.06$, 17nN applied force -- dataset from Fig. S10, cyan). Analysis of the particle rotation during the forward and backward displacement, respectively corresponding to positive and negative values of $\Delta \theta_{x}$, shows how, for rough, non-adhesive surfaces, $\Delta \theta_z$ can be up to 5 times larger than the principal rotations $\Delta \theta_x$ at the corresponding time points.
Additionally, the values of $\Delta \theta_z$ are broadly scattered and can be both positive and negative, irrespective of the direction of principal rolling, which we ascribe to the randomness of the process of individual asperities catching on each other, with spinning due to the interlocking of off-center asperities. 

Correspondingly, for the case of an adhesive flat surface (T = 36 \textdegree C, 18nN applied normal force -- dataset from Fig. \ref{fig4}A, orange), minor rotations (spinning around $z-$axis) are less pronounced but still present. $\Delta \theta_z$ is generally smaller than $\Delta\theta_x$ and rolling is much more regular too, as shown by the tight cloud of orange points.
The regularity of adhesion-induced rolling in comparison to roughness-induced rolling is further emphasized by the histograms of $\Delta\theta_x$ and $\Delta\theta_z$ for the two surfaces (Figs. \ref{fig7}B, C).
Note that, in both cases, the minor rotations are centered around $\Delta\theta_z$ = 0 \textdegree, suggesting that there is no bias in either direction and that spinning around the z-axis is random. 

We would like to remark that the ability to visualize minor rotations is unique to our methodology. In fact, in previous studies using sharp AFM tips for particle pushing and measuring rolling friction \cite{Ma2020FrictionManipulation, Sumer2008RollingPushing, Sitti2004AtomicCharacterization} any off-center forces, which might give rise to spinning, would cause particles to move sideways off the pushing tip, and must therefore be avoided. By contrast, the design of our probe is "self-centering", and avoids this problem, enabling a more comprehensive characterization of rolling in all directions.

\section{CONCLUSIONS}
Our results show that only considering sliding in the description of interparticle contacts in dense suspensions constitutes a severe limitation. Contacts between free particles clearly allow for rotation in all directions, depending on the detailed contact geometry and interparticle forces. The experimental approach that we present here advantageously allows for screening the effect of a wide range of additional particle features and environmental parameters, such as particle size, or pH and salt concentration, which we envisage exploring in the near future. 
Moreover, we demonstrated that rolling is associated with lower friction and hence less dissipation at contact. Correspondingly, particles would preferentially tend to roll in the presence of enough traction, unless forced to slide. 
This last consideration has strong implications for force transmission in dense particle networks, such as the ones emerging in ST suspensions \cite{Nabizadeh2022StructureSuspensions}. Our experiments in fact only mimic pair-wise contacts between two particles, while the onset of rigidity in DST systems has been associated with the percolation of four-particle clusters in contact \cite{Goyal2024FlowSuspensions}. If each particle is in contact with three neighbors, not only does this provide rigidity against deformation, but it also blocks rotation around all axes. Consequently, in the cases where blocking rolling leads to a substantial increase in effective friction, e.g. as shown here for rough and adhesive particles, we can expect stronger DST.
The fact that rough surfaces and asperity interlocking promote rotations around all axes, while smooth, adhesive particles tend to roll along the principal direction, suggests that the spatial distribution of force fluctuations in DST suspensions of these two particle types are likely to be very different.  
Additionally, because rolling and sliding contacts have different durations, we also expect the temporal fluctuations of the transmitted forces to be strongly affected by the relative particle motion. In dense suspensions, there will be a mixture of sliding and rolling contacts depending on the local microstructure and forces, and we expect sliding contacts to contribute to faster force fluctuations than rolling ones. Previous work has shown that adding small fractions of particles with different surface properties leads to dramatic changes in DST \cite{Hsu2018Roughness-dependentThickening, AkbariFakhrabadi2021AlteringSilica}, and we envision that ''additives'' that specifically promote or restrict rolling or sliding will have tremendous potential. These considerations shed new light on the design of particle systems with tailored surfaces for smart rheological formulations \cite{Jackson2022DesigningChemistry}.
Finally, we envisage that the method reported in this manuscript will enable the direct measurement of rolling friction coefficients for a broader range of systems, which can then be used as input for detailed numerical simulations, where the spatial distribution and the dynamical evolution of the contact network can be fully resolved during DST.

\section{MATERIALS AND METHODS}
All chemicals, if not stated otherwise, were used as provided by the supplier. Ethanol (EtOH, 96 \%), propylene glycol monomethyl ether acetate (PGMEA, 99.5 \%), 2-propanol (IPA, 99.5 \%), glucose (99.5 \%), ammonia (25 \% in water), hydrogen peroxide solution (30 \% in water), poly(diallyldimethylammonium chloride) solution (PolyDADMAC, 20 \%w/v in water, 400-500 kDa), tetraethyl orthosilicate (TEOS, 99 \%), polyethylenimine (PEI, M$_{n}$ = ~10,000), (3-aminopropyl)triethoxysilane (APTES, 97 \%), (3-mercaptopropyl)trimethoxysilane (MPTS, 95 \%), chloroform (CHCl$_3$, 99.8 \%), CdSe/ZnS core-shell type quantum dots (QD, stabilized with octadecylamine ligands, $\lambda_{em}$ = 520 nm, solid), dichloromethane (DCM, 99.8 \% anhydrous), $\alpha$-bromoisobutyryl bromide (BIBB, 99 \%), triethylamine (TEA, 99.5 \%), copper(II)bromide (CuBr$_{2}$, 99 \%), tetraethylammonium bromide (TEAB, 98 \%) and tris[2-(dimethylamino)ethyl]amine (Me$_{6}$TREN, 97 \%) were purchased from Sigma-Aldrich (Switzerland). N-isopropylacrylamide (NIPAM, 97 \%, Sigma-Aldrich, Switzerland) was recrystallized from toluene/hexane (3:2) and dried under vacuum before use. Copper(I)bromide (CuBr, 99.99 \%, Sigma-Aldrich, Switzerland) was purified by stirring in glacial acetic acid for 12 h before filtering,  washing several times with acetone and diethyl ether and finally drying in a vacuum.
\paragraph*{Probe Fabrication} The 3D-model of the holder was designed in AutoCAD2022 and exported as a .stl-file. The dimensions and design features of the holder are illustrated in Fig. S1. A fused-silica substrate (Multi-Dill, NanoScribe GmbH, Germany) was prepared according to the standard procedure from Nanoscribe (rinsed with EtOH, plasma-treated for 20 s with Piezobrush PZ2, relyon plasma GmbH, Germany). A commercial 2-photon-polymerization, direct-laser-writing setup (Photonic Professional GT2, NanoScribe GmbH, Germany) with a 63$\times$, NA = 1.5 objective and a commercial negative photoresist (IP-Dip2, NanoScribe GmbH, Germany) were used to fabricate arrays of 20-30 holders using the standard printing recipe for IP-Dip2. The holders were printed with the cavity facing up to avoid trapping uncured resin in the next steps. The finished prints were developed using the standard procedure from Nanoscribe (20 min PGMEA, 5 min IPA, dried in nitrogen stream) and post-cured for 1 h using a UV-light (365 nm). A microscope glass slide (Thermo Scientific, Switzerland) was plasma-treated for 20 s before a glucose layer (40 w\% in MilliQ water) was spin-coated (Laurell Technologies Corp., USA) on it at 4000 rpm for 15 s. The substrates with the holder array and with the sacrificial glucose layer were carefully brought into contact without applying any additional force. When separated, the holders transferred to the substrate with glucose, turning them upside down in the process. Once transferred, the holders could be stored until their final assembly. Holders were attached to tipless AFM cantilevers (HQ:CSC38/tipless/Cr-Au, MikroMasch, Bulgaria) to complete the fabrication of the free colloidal probe. The cantilever and the glucose substrate with holders were UV-ozone cleaned (Ossila Ltd., UK) for 10 min. After mounting the clean cantilever on an AFM (Dimension Icon, Bruker, USA), a small amount of UV glue (Norland Optical Adhesive 81, USA) was placed at its end by touching the edge of a drop. It was then aligned with the center of one of the holders and brought into contact at 10 nN applied normal force. The UV-adhesive was then cured by exposing it to UV light (GEM10 UV, 365 nm, Nitecore, Germany) for 1 min. The glucose layer was softened by increasing the local humidity with a wet paper towel to release the holder from the substrate. The assembled probe was submerged in MilliQ water for 10 min to remove any glucose residue on the holder. The final alignment and integrity of the probe were confirmed with optical microscopy (BX41, Olympus, Japan) and scanning electron microscopy (Gemini Leo-1530, Germany). Finally, the normal and torsional resonance frequency and quality factor of the cantilevers with the holders were determined in air with an AFM (Nanowizard III, JPK, Germany).
\paragraph*{Synthesis of smooth particles with fluorescent markers} 
The synthesis of quantum-dot (QD) labeled silica particles was inspired by previous work \cite{Jo2020SensitiveBeads, Jun2012UltrasensitiveBioimaging, Xie2020HighlyDevices, Hahm2023SilicaTrimethoxy2-Phenylethylsilane}. Silica core particles (12 $\mu$m, microParticles GmbH, Germany) were cleaned by mixing 1 mL ammonia (25 \% in water) with 1 mL H$_{2}$O$_{2}$ (30 \% in water) and 0.8 mL MilliQ water and heating the mixture to 70 \textdegree C. 0.2 mL of the silica particles (5 wt\%) were added and stirred for 10 min before retrieving the particles by repeated centrifugation in MilliQ water and EtOH. 10 mg of cleaned silica particles in 1 mL EtOH were added to a flask containing 2 mL EtOH, 10 $\mu$L MPTS, and 25 $\mu$L ammonia (25 \% in water) and stirred for 12 h. The particles were washed by repeated centrifugation in EtOH and finally dispersed in 2 mL EtOH. 5 $\mu$L of QD in CHCl$_3$ (10 mg/mL) were added to the suspension and shaken for 1 min before adding 10 mL CHCl$_3$ and shaking for 5 min. Finally, 10 mL CHCl$_3$, 0.1 mL MPTS, and 0.1 mL ammonia (25 \% in water) were added and stirred for 1 h. The resulting suspension was washed by repeated centrifugation in EtOH. A silica shell was grown on the RB particles for a chemically homogeneous surface and to bind the QD to the core. To this end, 6.5 mL EtOH, 1.3 mL ammonia, and 1 mL MilliQ water were added to the particles, suspended in 1 mL EtOH. A 5-vol\% solution of TEOS in EtOH was added sequentially. First, 0.25 mL of the TEOS solution was added at 2 mL/h (NE-1000, New Era Pump Systems Inc., USA) while stirring. After 30 min of stirring, the process was repeated before adding another 0.25 mL at 2 mL/h while sonicating (SONOREX SUPER RK 100 H, Bandelin, Germany) the solution. The particles were cleaned by repeated centrifugation in MilliQ water and analyzed using scanning electron microscopy (Gemini Leo-1530, Germany).
\paragraph*{Synthesis of raspberry-like particles with fluorescent markers} The synthesis of raspberry-like RB particles was reported previously \cite{Zanini2017FabricationHeteroaggregation,Ilhan2020SphericalDimensions} and was adapted for larger RB particles. Silica core particles (12 $\mu$m, microParticles GmbH, Germany) were cleaned by mixing 1 mL ammonia (25 \% in water) with 1 mL H$_{2}$O$_{2}$ (30 \% in water) and 0.8 mL MilliQ water and heating the mixture to 70 \textdegree C. 0.2 mL of the silica particles (5 wt\%) were added and stirred for 10 min before retrieving the particles by repeated centrifugation in MilliQ water. To invert the surface charge of the particles, they were added to 10 mL of an aqueous solution of PolyDADMAC (0.025 wt\%) and stirred for 1 h. The particles were washed by repeated centrifugation in MilliQ water. A batch of RB particles was produced using the same size of fluorescent-tracker berries and silica berries. The berry particles were attached via electrostatic aggregation by mixing the core particles in 10 mL MilliQ water with fluorescent polystyrene particles (100 nm: 0.875 mL of a 0.005 \%w/v suspension from Invitrogen, USA, with abs/em = 505/515 nm, 300 nm: 1.76 $\mu$L of a 2.5 \%w/v suspension from microParticles GmbH, Germany with abs/em = 502/518 nm) while stirring for 20 min, followed by silica particles (100 nm: 20 $\mu$L of a 1 \%w/v suspension from nanoComposix, USA, 300 nm: 40 $\mu$L of a 1 \%w/v suspension from nanoComposix, USA) for another 80 min. The RB particles were left to sediment and the supernatant with the remaining berries was removed to concentrate the dispersion to 1 mL. For a chemically homogeneous surface and to bind the berries to the core, a silica shell was grown on the RB particles. 7.5 mL EtOH and 1.3 mL ammonia were added to the RB particle suspension. A 5-vol\% solution of TEOS in EtOH was added sequentially. First, 0.25 mL of the TEOS solution was added at 2 mL/h (NE-1000, New Era Pump Systems Inc., USA) while stirring. After 30 min of stirring, the process was repeated before finally adding another 0.25 mL at 2 mL/h while sonicating (SONOREX SUPER RK 100 H, Bandelin, Germany) the solution. The RB particles were cleaned by repeated centrifugation in MilliQ water and analyzed using scanning electron microscopy (Gemini Leo-1530, Germany) and atomic force microscopy (Dimension Icon, Bruker, Germany). 
\paragraph*{Synthesis of rough substrates} A glass cover slide (18$\times$18 mm, \#1, Menzel-Gläser, Germany) was cleaned by sonication in toluene, IPA, EtOH, and MilliQ water for 10 min each before drying with nitrogen and UV/ozone cleaning (Ossila Ltd., UK) for 20 min. The substrate was submerged in an aqueous solution of PEI (1 mg/mL) for 30 min while stirring. After rinsing with MilliQ water and drying with nitrogen, the cover slide was placed in an aqueous suspension of silica particles. The coverage of the particles was determined by their concentration in the suspension and how long the substrate was submerged. Suspensions with 0.0025 wt\% of 50 nm particles (nanoComposix, USA), 0.04 wt\% of 100 nm particles (nanoComposix, USA), 0.02 wt\% of 200 nm particles (nanoComposix, USA), 0.04 wt\% of 300 nm particles (nanoComposix, USA), 0.04 wt\% of 400 nm particles (microParticles GmbH, Germany) and 0.08 wt\% of 500 nm particles (microParticles GmbH, Germany) were prepared. Substrates were submerged for 20 min (100 nm), 20 min (200 nm), 60 min (300 nm), 150 min (400 nm), and 35 min (500 nm), respectively, to prepare substrates with particles of different sizes. Four PEI-functionalized slides were submerged in the 50-nm-particle suspension and retrieved after 5, 30, 50, and 80 min to prepare substrates of different particle coverages. All substrates were carefully rinsed with MilliQ water and finally submerged in a TEOS solution (1 vol\% in EtOH) to permanently bond the particles to the cover slide. The substrates were submerged in a container with 7.44 mL EtOH, 1.22 mL ammonia, and 1 mL MilliQ water before adding 0.6 mL TEOS solution and stirring for 30 min. After rinsing with MilliQ water and drying in a nitrogen jet, the substrates were analyzed using scanning electron microscopy (Gemini Leo-1530, Germany) and AFM (Dimension Icon, Bruker, Germany). 
\paragraph*{Characterisation of RB particles and rough substrates} A microscope glass slide (Thermo Scientific, Switzerland) was submerged in a solution of PEI (1 mg/mL) for 30 min before rinsing with MilliQ water and drying in a nitrogen stream. 10 $\mu$L of the raspberry dispersion were added to the glass slide. Once the water evaporated, AFM scans of single RB particles were obtained in tapping mode (Dimension Icon, Bruker, USA, OMCL-AC160TS, Olympus, Japan) to analyze the surface roughness using a custom-written Matlab code (Fig. S8). A single RB particle was centered in the frame and the global curvature, based on the RB particle radius, was subtracted to obtain the true surface topography of the RB particle, decoupled from the shape of the RB particle. Finally, all asperities were identified by finding the local maxima. The average height of all asperities (\textit{h}) and the average distance to the nearest neighbors (\textit{d}) were recorded to calculate the dimensionless roughness parameter \textit{h/d}. The rough substrates were UV/ozone cleaned for 10 min before obtaining tapping mode AFM scans. The same roughness analysis to determine \textit{h/d} was performed, see Fig. S6 and S7. 
\paragraph*{Immobilization of SI-ATRP initiator for subsequent PNIPAM functionalization of surfaces} First, the freshly synthesized particles (i.e., smooth particles labeled by QDs and rough RB particles) were functionalized with APTES. 50 mg particles were dispersed in 10 mL dry EtOH before adding 0.1 mL APTES and stirring for 12 h in a nitrogen atmosphere. The functionalized particles were cleaned three times by centrifugation in EtOH and dried under a vacuum. The particles were redispersed in 50 mL dry DCM and degassed with nitrogen for 30 min. Next, the particles were reacted with 0.2 mL BIBB and 0.4 mL TEA in a nitrogen atmosphere for 4 h before cleaning them by centrifugation in DCM three times and drying them in a vacuum.
\paragraph*{Synthesis of PNIPAM brushes on particles} 50 mg of the dry initiator-grafted particles were dispersed in 2.6 mL of an EtOH/water (4:1) mixture and degassed with nitrogen for 1 h. Next, 2.4 g NIPAM, 0.41 g TEAB and 19.2 $\mu$L Me$_{6}$TREN were dissolved in 5 mL EtOH/water (4:1) and degassed with nitrogen for 1 h. The solution was then transferred with a degassed syringe to a flask containing 6.9 mg CuBr and 5.4 mg CuBr$_{2}$ in nitrogen and stirred until the catalyst was dissolved. The solution was transferred with a degassed syringe to the particle solution to start the polymerization. After 30 min, the reaction was quenched by adding 10 mL EtOH/water (4:1) mixture and exposing the suspension to air. The PNIPAM-functionalized particles were cleaned by repeated centrifugation in MilliQ water and stored in MilliQ water until further use. The PNIPAM particles were analyzed using scanning electron microscopy (Gemini Leo-1530, Germany).
\paragraph*{Synthesis of PNIPAM brushes on smooth and rough glass slides} Substrates were functionalized with PNIPAM in the same fashion as the PNIPAM particles. First, smooth glass cover slides (18$\times$18 mm, \#1, Menzel-Gläser, Germany) and rough substrates (100 nm silica nanoparticles) were functionalized with APTES in dry EtOH before rinsing with EtOH and drying with a nitrogen jet. Next, the substrates were submerged in a solution of BIBB and TEA in dry DCM after degassing with nitrogen. After 4 h the substrates are washed with DCM and dried with a nitrogen jet. The grafting of the PNIPAM brushes was carried out using the reactant equivalents of 50 mg PNIPAM particles per substrate and a reaction time of 30 min. The substrates were cleaned with toluene, EtOH, and MilliQ water and dried with a nitrogen jet. The thickness of the PNIPAM brush was determined to be 16 nm using ellipsometry (M-2000 Ellipsometer, J.A. Woollam Co., USA) 
\paragraph*{Experimental procedure for the free colloidal probe} To perform simultaneous friction- and rotation-tracking experiments, a combination of an AFM (Nanowizard III, JPK, Germany) and an inverted microscope (Axio Observer D1, 40$\times$ NA = 0.6 objective, Filter Set 09, Zeiss, Germany) with fluorescent imaging capabilities (89 North Photofluor II, USA) was required. The probe and the substrates (except for  PNIPAM functionalized substrates) were UV cleaned for 10 min before use. A dilute suspension of particles with fluorescent markers in 10 mM aqueous HEPES buffer (pH = 7.4) was added to the substrate and mounted on the stage of the microscope. The probe, mounted to the AFM holder was submerged in the buffer droplet. After the standard AFM laser-alignment procedure, a single particle was located on the substrate, the holder was positioned above and the particle was captured by approaching the substrate. The sensitivity of the cantilever was determined by performing force-distance curves at the beginning of every experiment with an approach/retraction speed of 500 nm/s. Next, friction loops were acquired at user-defined applied normal forces (ranging from 2 to 60 nN) while simultaneously obtaining time-series images. The scan speed and scan size were set to 10 $\mu$m/s and 50 $\mu$m, respectively. 10 friction loops were obtained for every particle at each applied normal force. When lifting the probe from the substrate, the particle was released. To account for particle and substrate morphology variations, several particles were measured in different locations and the average friction force and rotation were determined. 
\paragraph*{Analysis of the 3D rotational motion from 2D images} The rotation of the optically anisotropic particles was extracted using a method developed by Niggel et al. \cite{Niggel20233-DSurfacesb} and briefly illustrated in Fig. S4. A custom Matlab code was used to track the particle in a time series of images recorded during the friction experiments. The rotation was then extracted from cropped images of the centered RB particle. The image of the particle at time \textit{t} was projected onto a sphere and the sphere was virtually rotated by user-defined angles. The rotated image was then compared to an image at time \textit{t + dt} to find the most probable change in angular orientation. The threshold for rolling was determined from the noise level in the rotational tracking of a stationary particle (pinned down by the probe, but not translated) and set to $\pm$ 1\textdegree\ in most cases. The instantaneous rotation was synchronized with the lateral force to determine local correlations between the friction and motion of a particle, as seen in Fig. \ref{fig1}. The cumulative rotation, compared to the theoretical rotation of a particle without slip, illustrates the average ratio of rolling vs. sliding of a particle, as seen in Fig. \ref{fig3} and Fig. \ref{fig4}.
\paragraph*{Analysis of the lateral-force signal} The analysis of the lateral-force signal was based on the standard procedure to extract the friction coefficient from colloidal-probe atomic force microscopy in lateral force mode. A custom Matlab code was employed to convert the raw voltage signal to the lateral force using a calibration factor and extract the average lateral force for a certain applied normal force. The sliding friction coefficient $\mu$ was extracted from the slope of a linear fit to the plot of average lateral force versus normal force for fixed particles and free particles with $\theta_{x}/\theta_\mathrm{th}$ $<$ 50 \%.
For friction datasets of free particles with average $\theta_{x}/\theta_\mathrm{th}$ $>$ 50 \%, we calculated the rolling friction coefficient $\eta/R_p$, following the analysis outlined in the \textit{Results}. In particular, the sliding friction of the particle against the holder $\mu_1$ was determined from friction loops at the onset of rolling ($\theta_{x}/\theta_\mathrm{th}$ = 30 - 70 \%) for each particle -- holder combination and used to obtain $M/R_p$. Finally, the rolling friction coefficient $\eta/R_p$ was calculated from the slope of the linear fit of the positive values of $M/R_p$ versus $F_N$.
\paragraph*{Conversion of the free colloidal probe to a fixed colloidal probe} A particle could be irreversibly attached to the holder, restricting any rotation and turning it into a sliding friction probe. In that way, a fixed and free RB particle could be directly compared using the same cantilever. The probe was mounted on an AFM (Nanowizard III, JPK, Germany), and a glass microscope slide (Thermo Scientific, Switzerland) with dried particles was placed on the sample stage together with a small drop of UV-adhesive (Norland Optical Adhesive 63, USA). The holder was briefly dipped into the glue before approaching the flat substrate several times to remove any adhesive from the bottom of the holder. Next, a particle was located and centered before approaching it with the probe to capture it. The glue was hardened using a UV flashlight (GEM10 UV, 365 nm, Nitecore, Germany) for 1 min as soon as the holder was in contact with the particle. 
\paragraph*{Calibration of the free colloidal probe} The raw signal of the vertical and torsional bending of the cantilever was obtained as the change in deflection of a laser beam that was reflected off the back of the cantilever onto a position-sensitive detector (PSD). This voltage signal was converted to force via two separate calibration methods for the normal and lateral forces acting on the cantilever. For the normal force, the established Sader method \cite{Sader1999CalibrationCantilevers} was used, which is based on the experimentally determined, plan-view cantilever dimensions, the resonance frequency, and the quality factor. Additionally, the sensitivity of the cantilever was determined from force-distance curves obtained at the start of every experiment. The lateral force calibration was performed using the improved wedge calibration method by Varenberg et al. \cite{Varenberg2003AnMicroscopy}. The calibration factor $\alpha$ was determined from the change in lateral deflection as the cantilever was scanned over an area with two well-defined slope angles (see Fig. S2).

\begin{acknowledgments}
The authors thank Jan Vermant and Lars Pastewka for inspiring discussions, André R. Studart for providing access to his SEM facilities, and acknowledge the financial support of ETH Zurich.
\end{acknowledgments}


\end{document}